# End-to-end topographic networks as models of cortical map formation and human visual behaviour: moving beyond convolutions

Zejin Lu*[1,4], Adrien Doerig*[1], Victoria Bosch*[1], Bas Krahmer,
Daniel Kaiser[+2,3], Radoslaw M Cichy[+4], & Tim C Kietzmann[+1]

[1] Machine Learning, Institute for Cognitive Science, Osnabrück University, Osnabrück, Germany.
[2] Neural Computation Group, Mathematical Institute, Justus-Liebig-Universität Gießen, Gießen, Germany.
[3] Center for Mind Brain and Behavior, Philipps-Universität Marburg and Justus-Liebig-Universität Gießen, Marburg, Germany
[4] Neural Dynamics of Visual Cognition Group, Department of Education and Psychology, Freie Universität Berlin, Berlin, Germany.
* Shared first authorship.
+ Shared last authorship.

## Abstract

Computational models are an essential tool for understanding the origin and functions of the topographic organisation of the primate visual system. Yet, vision is most commonly modelled by convolutional neural networks that ignore topography by learning identical features across space. Here, we overcome this limitation by developing *All-Topographic Neural Networks* (All-TNNs). Trained on visual input, several features of primate topography emerge in All-TNNs: smooth orientation maps and cortical magnification in their first layer, and category-selective areas in their final layer. In addition, we introduce a novel dataset of human spatial biases in object recognition, which enables us to directly link models to behaviour. We demonstrate that All-TNNs significantly better align with human behaviour than previous state-of-the-art convolutional models due to their topographic nature. All-TNNs thereby mark an important step forward in understanding the spatial organisation of the visual brain and how it mediates visual behaviour.

# Introduction

Artificial neural networks (ANNs) have enabled the investigation of neuroscientific questions that were previously beyond the scope of traditional modelling and experimental techniques by offering a way to design models that are image-computable, and task-performing, while bridging levels of explanation from single neurons to behaviour[1–3]. In vision, the most commonly used networks are convolutional neural networks (CNNs), a powerful and efficient architecture type that has been successful at predicting primate neural activity across multiple hierarchical levels of the ventral stream[4–7] and at accounting for complex visual behaviour[8–11].

However, a crucial limitation on the future prospects of CNNs as neuroscientific models is the architecture's reliance on weight sharing, i.e. CNNs use identical features across visual space. This strong inductive bias is sensible for engineering purposes, because it facilitates efficient learning and enables spatially invariant object recognition. However, this architectural design choice limits their ability to model fundamental aspects of biological vision. A central aspect is the origin and function of cortical topography[12,13] and its relation to behaviour - a central area of research in visual neuroscience for which modelling promises important insights that cannot easily be addressed using only experimental approaches.

In the brain, topographic organisation refers to the fact that the spatial arrangement of neurons on the cortical sheet is highly structured with respect to their tuning profiles. For example, early visual cortex is thought to be organised into columnar structures (hypercolumns) with repeating motifs of orientation sensitivity that vary smoothly across the surface[14–16]. In higher-level visual cortex, clusters of neurons that respond preferentially to abstract stimulus categories, such as faces[17,18], bodies[19], and scenes[20–22] are observed, among other spatial organisational structures based on a variety of visual and conceptual stimulus properties[23–25]. Human visual behaviour, too, exhibits spatial regularities, with objects being more easily recognized when displayed in their typical spatial position[26–28], likely arising from the topographic organisation of visual cortex with joint spatial tuning and feature tuning determining visual efficiency.

The emergence of topographies and its interrelation with behaviour cannot directly be modelled using CNNs due to their spatially enforced weight sharing, leading to three modelling limitations First, synaptic changes in CNNs are globally orchestrated to form identical feature selectivity across space, which is in stark contrast to the brain where synaptic changes across the cortical topography are local. Second, CNNs lack a clear spatial arrangement similar to the brain's cortical sheet. Third, CNNs do not exhibit spatially smooth neural tuning transitions as found in the brain. Here we overcome these limitations with a new topographic model architecture, which we term *All-Topographic Neural Network (All-TNN).* All-TNNs fulfil the following desiderata for modelling cortical topographic organisation:

1) *Locality:* Units in the model need to have local receptive fields (RFs) that are learnt individually and not enforced to be exact duplicates of other RFs.
2) *Arrangement along the cortical sheet:* The spatial smoothness across cortex is thought to be due to a smooth decay in connectivity between neurons with increasing cortical distance[29,30]. Units in the model, therefore, need to be arranged along an artificial cortical sheet, and the spatial distance between units is to be measured in this space.

3) *Smoothness:* There is a continuum from spatially discontinuous models where each unit detects a different feature, to spatially uniform models where all units detect the same feature. A biologically plausible model of topographic organisation should reflect the biological observation that the brain operates in between these two extremes[25,31–33].

Across a set of experiments with this architecture and a novel dataset of spatial biases in human visual object recognition, we demonstrate that All-TNNs, in contrast to current state-of-the-art CNNs, more accurately capture topographic representations in the primate visual system. First, they reproduce important properties (smooth orientation selectivity maps, cortical magnification and category-selective regions) of neural topography when trained on visual input. Second, All-TNNs significantly better align with human behaviour by reproducing visual field biases in object perception. Lastly, we show that the visual behaviour of All-TNNs is directly linked to their topography.

## Results

To study the emergence of feature topographies and their impact on behaviour, we developed a fully topographic neural network, All-TNN. Contrary to CNNs, All-TNNs can learn locally specific weight kernels to detect different features across visual space (Fig. 1, desideratum 1). As a result, units at different spatial locations are free to learn different features, making it possible to directly compare the topographies of spatial selectivity maps on the model's cortical sheets with properties of topographies found in the brain. In addition, units of each network layer are arranged along a 2D cortical sheet (desideratum 2), while resembling hypercolumnar structure: all units of a hypercolumn share the same local receptive field location (i.e., they only receive connections from the same spatially limited area in the layer below). To navigate the continuum from spatially discontinuous to spatially uniform feature selectivity, we use a tunable spatial similarity loss that acts as a regularizer encouraging neighbouring units to detect similar features (desideratum 3; see Methods).

In our experiments, we trained All-TNNs on ecoset, an object classification dataset that contains 565 object categories selected to be representative of concrete categories that are of importance to humans[34]. Training for object categorisation performance while satisfying the spatial similarity loss results in a dual-loss objective, which forces the network to trade off between i) learning varied feature selectivity required to accomplish a difficult object categorisation task, and ii) preserving similar feature selectivity between neighbouring units. In our experiments, we contrasted multiple instances of All-TNN with two control models: purely locally connected networks (LCN, i.e. All-TNNs without spatial similarity loss), and CNNs. To make sure we single out architectural differences in our analyses, these models have matching numbers of units, identical hyperparameter settings, and are trained on the same dataset and task (see Methods). We train multiple seeds of each network type that we treat as experimental subjects.

To confirm the capacity of All-TNN to model cortical topography and its relation with behaviour, we perform in-silico electrophysiology analyses hierarchically from low-level to high-level neural topographical characteristics, and then move onto behavioural experiments.

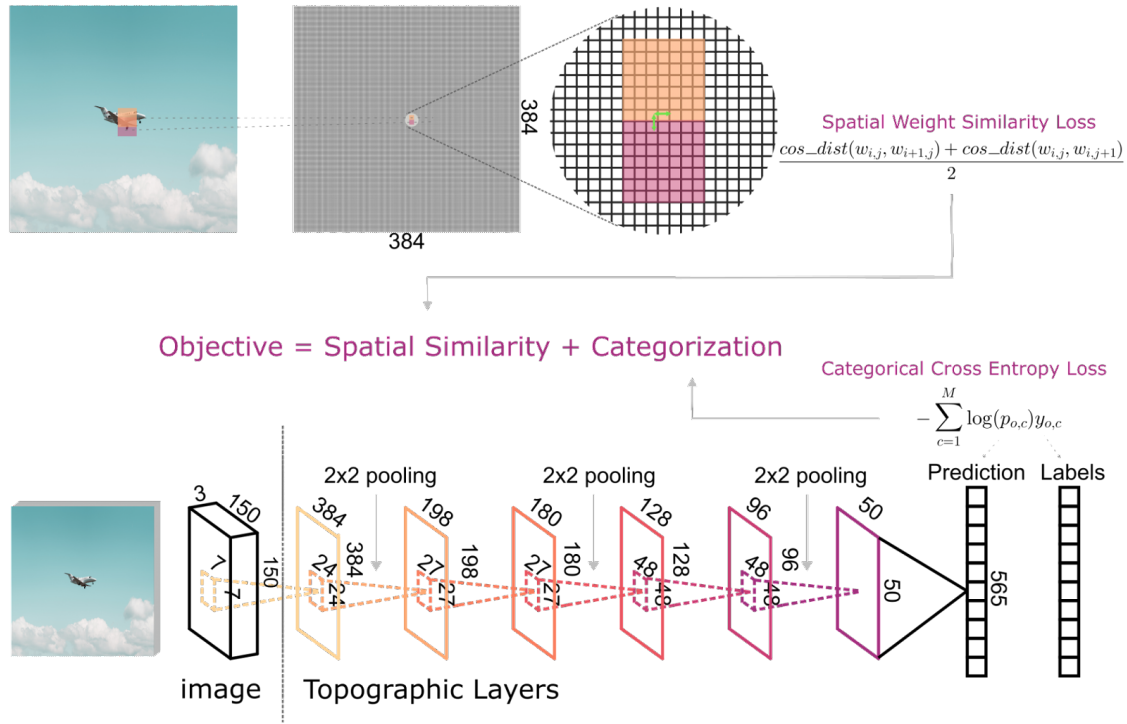

**Figure 1 | The All-Topographic Neural Network (All-TNN).** *Example topographic layer (top) with the properties of local connectivity, 2D arrangement and spatial weight similarity loss. Units are arranged retinotopically into 'hypercolumns', and units in a given hypercolumn share the same local receptive field location. The spatial similarity loss is applied to weight kernels of neighbouring units within a layer (as illustrated for one unit by green arrows). The All-TNN architecture (bottom) used in our experiments consists of 6 topographic layers of different dimensions (indicated by numbers along the layer depictions) and kernel sizes (indicated by the numbers along the kernel depiction), of which layers 1, 3 and 5 are followed by pooling layers, and the last layer is followed by a category readout (565 categories) with softmax. The network's learning objective is the sum of a classification loss and a loss favouring local spatial similarity between unit kernels.*

*Topographical features of the ventral stream emerge in All-TNNs*

V1 orientation selectivity maps are a topographical hallmark of the primate visual system[12]. We thus begin our investigation by determining whether the model's first layer reproduces the features of smooth orientation selectivity maps in V1. To determine orientation selectivity for each unit in the layer, we follow the standard analysis procedure in biological systems[14] by presenting the network sinewave gratings of different angles and phases (Fig 2a) and determining the angle for which each unit is most responsive (see Methods). We find that the first All-TNN layer exhibits a smooth distribution of orientation selectivities, mirroring orientation selectivity maps in primate V1 (see Fig. 2a, left panel). By contrast, CNN architectures do not exhibit such topography, because feature selectivity is, by definition, identical across all locations in the layer (Fig. 2a, bottom right panel). Importantly, V1-like selectivity maps also did not emerge in the locally connected control network, suggesting that such a topographical organisation does not emerge purely from learning to categorise natural objects under the influence of the autocorrelation of the input statistics (Fig. 2a). In addition, while V1-like feature selectivity maps require training (they are not present in the untrained network), they emerge early and remain stable after only a few epochs (Fig. S1a), even though performance keeps increasing after the topography has stabilized. This suggests that the topographical organisation emerges in the network quickly, with further training only finetuning selectivities within this topography rather than bringing about broad changes in the overall structure. This is in line with early maturation of topographical structures in visual cortex of infants, thought to provide scaffolding for functional selectivity[35,36].

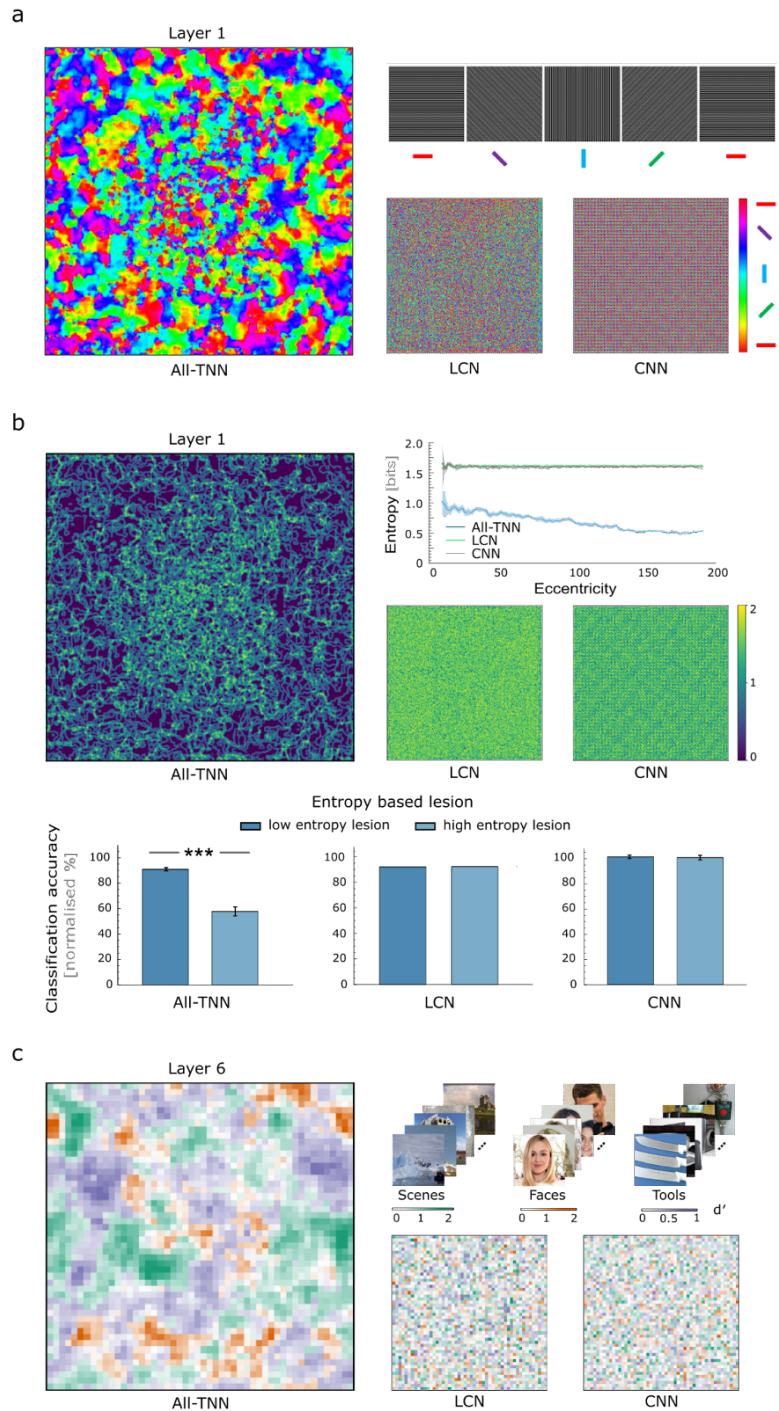

**Figure 2 | All-TNNs mirror key features of the visual system's topography.** *a. The first layer of All-TNN (example network instance) shows a V1-like organisation of orientation selectivities, while the two control architectures, a locally connected control network and a convolutional network, do not. b. Entropy visualisation of an All-TNN instance mirrors foveation and cortical magnification in the first layer. Entropy decreases with eccentricity in all seeds of All-TNN, but remains constant for CNN and LCN (data averaged across all seeds, shaded region shows the 95% confidence interval; curves overlap for LCN and CNN). Selective entropy-based lesioning confirms cortical magnification (data shown averaged*

*across all seeds, error bars show the variance). Classification accuracy is more affected when lesioning 50% of units in high-entropy (i.e. varied selectivity) regions of All-TNN, than lesions performed to units in low-entropy (i.e. homogeneous selectivity) regions. This effect of cortical magnification is neither observed for the locally connected control network nor the CNN. **c.** The last layer of All-TNNs shows clustering of high-level category-based selectivities (d') for tools, scenes, and faces, whereas the locally connected control network and the CNN do not show clustering of similarly selective units. Results and maps for all seeds can be viewed in Fig. S1-3.*

Interestingly, the All-TNNs' orientation selectivity maps exhibit a strong centre-periphery organisation, with increasingly smooth feature selectivity towards the periphery. To quantify the observation of a foveal region with a higher diversity of feature selectivities, we computed feature-entropy at various spatial eccentricities (Fig. 2b, top left panel), as well as for the control models (Fig. 2b, middle right panel). Indeed, we observe a marked decline in entropy in All-TNNs, in line with less varied feature selectivity in the periphery (Fig. 2b, top right panel). In contrast, entropy is consistently high and does not vary for control networks, which are not able to pick up on the image statistics in their topography. This aspect may be surprising since All-TNNs do not have a central bias in its architecture or loss terms. Instead, the position of this "foveal" region must therefore result from an interplay of the network's training objective with the statistics of the training dataset. We hypothesise that the centre of the images contains crucial information for the categorization task, which forces the network to learn varied features in this region at the expense of feature smoothness. In contrast, the network favours more homogenous visual features in the periphery, because less categorization-relevant information is present there.

This foveal bias is reminiscent of cortical magnification in humans, with more neurons per degree of visual angle in the foveal region of V1 than in the periphery, leading to better acuity in the fovea[37–39]. This allocation of more resources to the fovea goes hand in hand with the fact that humans fixate on relevant regions of the visual field through eye movements. We tested whether the greater diversity of foveal selectivities in All-TNNs reflects additional computational resources contributing to task performance, similar to human cortical magnification, in an in-silico lesion study. We quantified the diversity of selectivities at each location on the sheet by computing the local entropy of orientation selectivity in a sliding window. We then lesioned 50% of units in regions with homogenous selectivities (low entropy lesions) or varied selectivity (high entropy lesions) (Fig. 2b, bottom panel; see Methods). We found that low-entropy lesions have a minor effect on All-TNN classification performance (accuracy drop to 90.91% of the unlesioned All-TNN performance). In contrast, high-entropy lesions strongly deteriorate classification performance (accuracy drop to 57.87% compared to unlesioned All-TNN). This shows that All-TNNs can afford to lose units in the peripheral regions with homogenous selectivities, but not in the foveal region with diverse selectivities. Neither convolutional nor locally connected control models show this effect, because their selectivity profiles are homogeneous (Fig. 2b, middle right panel), in contrast to the strong topographical structure of All-TNNs. Hence, only All-TNNs can learn a topography mirroring the training image statistics to selectively process relevant parts of the visual field, leading to a spatial organisation reminiscent of cortical magnification.

Having investigated lower-level topographic features of All-TNNs, we focus on higher-level representations and analyse whether the networks' last layer reproduces topographical features characterizing primate higher-level visual cortex. To do so, we contrast unit activations in response to faces, tools, and places ($d'$, 500 stimuli each, see Methods) - image classes that yield clusters of neural activation in primate higher-level visual areas[17,18,20,21,23,36,40,41]. After model training, we observe smooth model regions selective for faces, places and tools (Fig. 2c, left panel) in the All-TNN. Neither of the control models shows a comparable topography. Instead, they develop an unstructured salt-and-pepper selectivity map without any clusters (Fig. 2c, bottom right panel). Similarly to the emergence of orientation selectivity in the first layer of All-TNN, this topographic organisation into category-selective regions requires training and stabilises early on (Fig. S3a).

Taken together, these results demonstrate that All-TNNs consistently reproduce key characteristics of primate neural topography at early and higher-level visual cortex, including V1-like smooth orientation maps, cortical magnification, and category-selective regions in the final model layer.

### All-TNNs better align with human spatial visual behaviour

Humans reliably show visual field biases, i.e. objects are better detected and recognized when they appear in the locations they are most often experienced in[26,27,42]. Given that, akin to cortical maps, All-TNNs have the ability to detect different features in different parts of the visual field, the question arises whether All-TNNs exhibit human-like effects of spatial position in their classification performance.

To investigate this question, we collected a novel dataset of spatial biases in human visual object recognition (Fig. 3a). Participants (n=30) classified 80 objects from 16 classes of the COCO dataset[43], which were presented for 40ms in a random location of a 5x5 grid, followed by a Mondrian mask (see Methods). Masking prevents ceiling effects in performance, and has been proposed to limit recurrent processing in humans[44], which is ideal as a testbed for our current set of feedforward models. Based on these data, we computed spatial classification accuracy maps for each individual and each object class that capture the spatial distribution of human classification performance (Fig. 3b).

We performed the same experiment on All-TNNs and our control models (see Methods; note that LCN control models are not considered further because they do not exhibit meaningful topographic organisation). Model testing was based on all images in COCO for a given class, instead of only 5 exemplars used for the human participants. In analogy to the human analyses, accuracy maps were computed for each model type, instance, and object category.

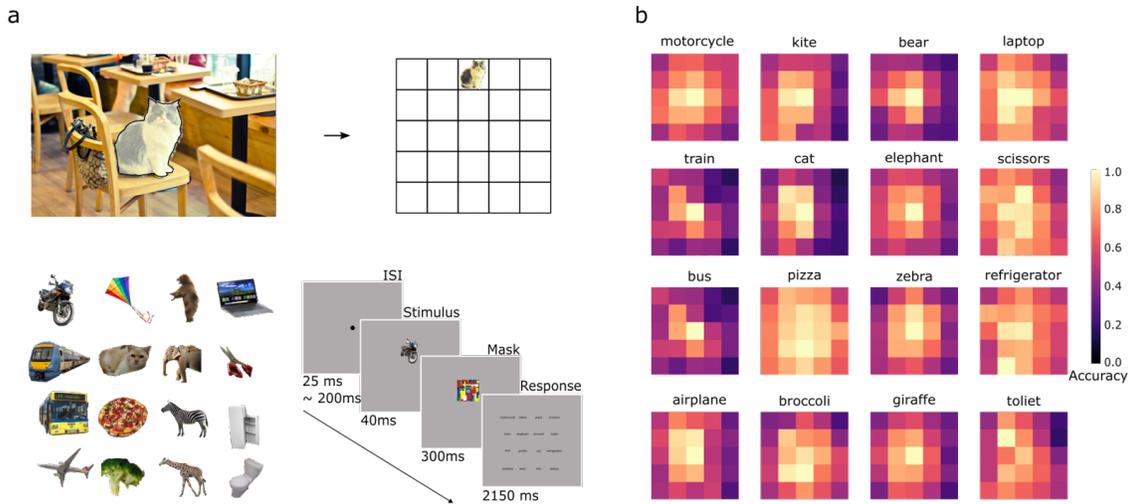

**Figure 3 | An experiment testing spatial biases in human visual behaviour.** *a. To create stimulus materials for the behavioural experiment, objects from COCO are segmented from their background and placed on a 5x5 grid. 16 object categories consisting of 5 object exemplars were included in this behavioural dataset (example segmentations shown). Each trial contained a brief stimulus presentation at one of 25 locations on the screen, followed by a Mondrian mask. A response screen showing the 16 target category labels was presented after the stimulus and mask to collect participant responses. **b.** Accuracy maps for all 16 categories, averaged across participants (n=30).*

As a first step to understanding spatial visual biases in both humans and models, we verified that object classification performance across the visual field is aligned with the corresponding object's occurrence statistics. To do so, we correlated the object occurrence frequency maps, as obtained from COCO (see Methods; Fig. S4), with the classification accuracy maps (Fig. 4a, left panel). For humans, we observe a positive relationship between accuracy maps and COCO occurrence frequency maps (Pearson r=0.56; Fig. 4a, right panel), consistent with previous research[26,27]. All-TNNs and CNNs, too, exhibit a significant correlation between accuracy and occurrence frequency (permutation test for both All-TNN and CNN, n=1e5; *p*<0.001). However, the strength of the effect observed in All-TNNs was significantly closer to humans compared to CNNs controls (Pearson's r; r=0.45 for All-TNN, r=0.23 for CNN, permutation test, n=1e5; *p*<0.001). This indicates that the alignment of performance and occurrence statistics is most similar to humans for All-TNNs than CNNs.

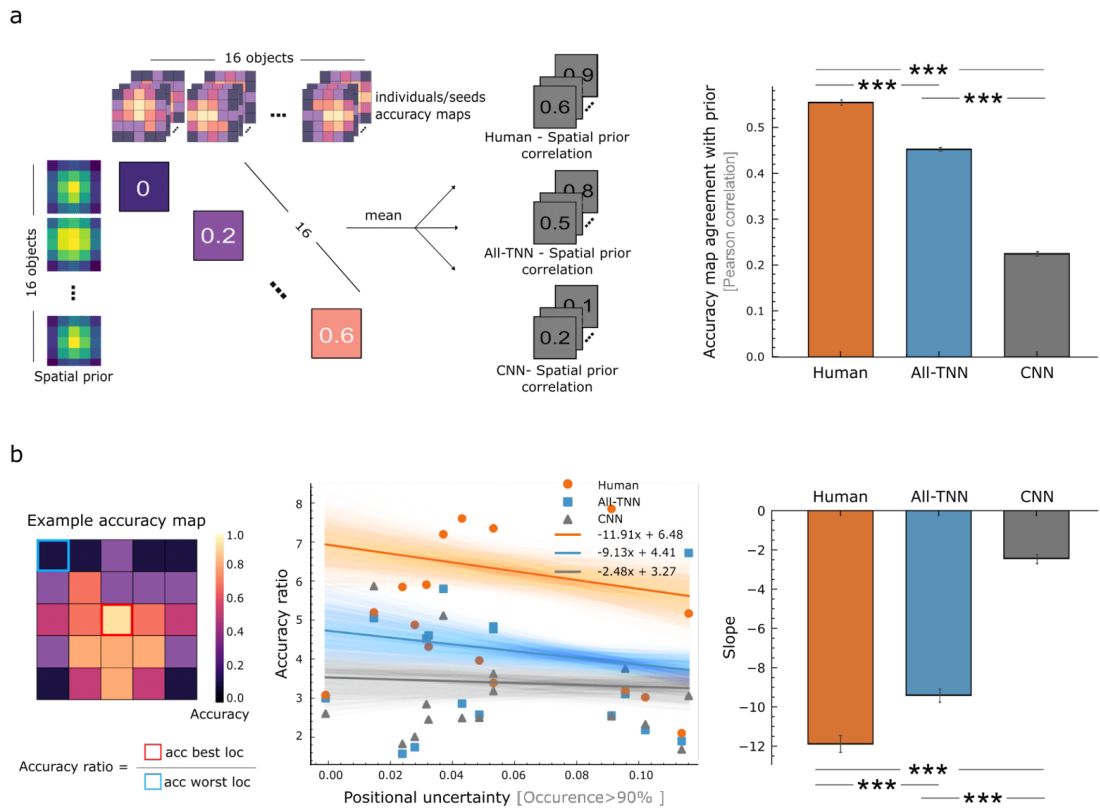

**Figure 4 | All-TNNs capture spatial statistics of objects, matching human behaviour.** *Data from human participants (n=30), All-TNN instances (n=10) and CNN controls (n=10) shown.* **a.** *Humans and All-TNNs show a stronger alignment (Pearson correlation) between their accuracy maps and occurrence frequency maps obtained from COCO for each of our selected 16 categories, as compared to CNNs.* **b.** *All-TNNs, similarly to humans, have less peaked accuracy maps when the location of objects is more variable. Positional uncertainty for each of our 16 categories was computed as the image area required to cover 90% of object occurrences in COCO. Positional variance in classification performance was computed by the accuracy ratio between the best and worst classification accuracy for a given object. Robust regression indicates a significant relationship between positional variance in classification performance and positional uncertainty across 16 categories for both humans and All-TNNs, but not CNNs. A negative slope indicates a decreasing accuracy ratio as a function of positional uncertainty.*

To further characterise positional effects and their relation to occurrence statistics, we investigated whether positional uncertainty of object categories in natural scenes had an effect on human and model accuracy maps. We hypothesised that object categories with stereotypical locations, i.e. with low positional uncertainty, should exhibit stronger behavioural differences across space due to stronger position-dependent tuning. For objects that occur in more diverse and unpredictable locations, however, behaviour should show weaker positional effects. We operationalise positional uncertainty as the size of the region where the object's occurrence frequency exceeds 90% of its maximum frequency, and positional effects on classification performance as the ratio between the locations with best and worst classification accuracy for each object class (Fig. 4b, left). We then tested for a relationship between these two measures via robust linear regression, and analysed the estimated slopes for average humans, All-TNNs, and CNN controls (Fig. 4b, middle and right). For human observers we observe a negative relationship: objects with low positional uncertainty exhibited stronger positional effects on accuracy (robust regression; avg. slope = -11.91; 95% CI, -22.56 - -0.18). Similar effects were also observed for All-TNNs (robust regression; avg. slope = -9.13; 95% CI, -18.86 - -0.40; permutation test, n=1e5; *p*<0.001). In contrast, CNN control models showed much-reduced effect sizes (robust regression; avg. slope = -2.48; 95% CI, -7.15 - 4.95; permutation test, n=1e5; *p*<0.001). These analyses indicate that the magnitude of the positional effect on classification performance varies as a function of how uniformly distributed object occurrences are, with All-TNNs again aligning more closely to human behavioural patterns than CNN control models.

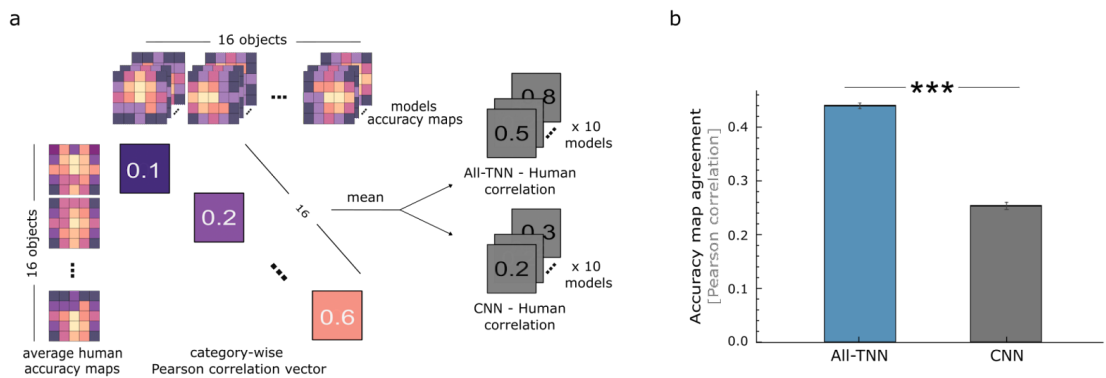

**Figure 5 | All-TNNs mirror spatial biases in human visual behaviour.** *a. Alignment of accuracy maps of humans and models was quantified by a Pearson correlation. b. All-TNNs exhibit significantly better alignment with human accuracy maps than CNNs. Shown are the noise-ceiling corrected agreements of All-TNNs and CNNs with human accuracy maps. The error bars show 95% confidence intervals.*

Having verified that humans and All-TNN models are able to mirror spatial occurrence statistics in their behavioural patterns, we next tested the models' ability to accurately predict human accuracy maps, i.e. variations in human object categorizations across the visual field. For this, we directly compared human behavioural and model accuracy maps through category-wise Pearson correlation analysis (Fig. 5a). We find that All-TNNs correlate significantly stronger with human behavioural patterns than CNN controls (*p*<0.001; n=10; Fig. 5b).

Could this pattern of results be caused by a simple centre bias instead of richer structure with different maps for different categories? To determine whether this is the case, or whether the behavioural alignment is more precise and indeed object specific, we constructed accuracy dissimilarity matrices (ADMs), by computing the Pearson correlation distance between all object-specific accuracy maps (Fig. 6a). This analysis is akin to representational similarity analysis[45], but based on our accuracy maps, highlighting in how far pairs of categories exhibit similar or different accuracy maps. ADMs are compared between humans and models via Spearman correlation. Correlating ADMs focuses on differences between accuracy maps. Indeed, features shared between accuracy maps, such as central biases, would show up as a constant in the ADM (all ADM cells are impacted by this one shared aspect). Constants, however, do not affect ADM comparisons using Spearman correlations. In short, our analysis approach of comparing ADMs focuses on differences in accuracy maps across object categories and thereby moves the analysis beyond similarities that are category-agnostic. As shown in Figure 6b, ADM agreement between human data and All-TNNs is significantly higher than between humans and CNN controls (permutation test, n=1e5; *p*<0.01). This confirms that visual classification behaviour of All-TNNs aligns with human behaviour in a category-specific manner rather than merely reflecting a central bias effect.

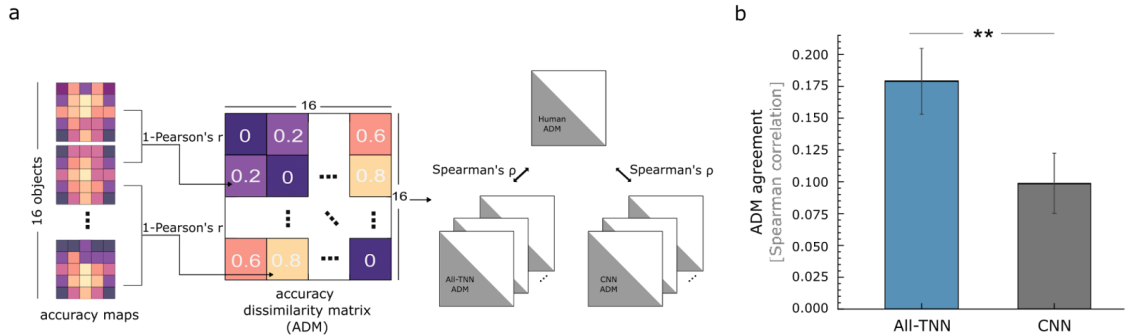

**Figure 6 | All-TNNs mirror object specific spatial biases in human visual behaviour.** *a. Accuracy dissimilarity matrices (ADMs) were created to capture the differences between accuracy maps of all objects using Pearson correlation distance (left). Behaviourally more similar objects have a low dissimilarity in their accuracy maps, whereas objects yielding behaviourally distinct accuracy maps have high dissimilarity. To relate the ADMs of average humans, All-TNNs, and CNNs with each other they are correlated using Spearman correlation (right).* **b.** *ADMs of All-TNNs align significantly better with human data than those of CNNs.*

*Engagement of stereotypical unit activation patterns links topography to behaviour*

When trained on natural scenes, All-TNNs develop units that are selective for different categories in different spatial locations. The behavioural experiment, however, relied on small cropped objects presented in different locations - a setting quite different from the natural images of ecoset (Fig. 7a, top panel). This discrepancy offers the unique opportunity to directly link the human-like behaviour of All-TNNs to their topographical arrangement.

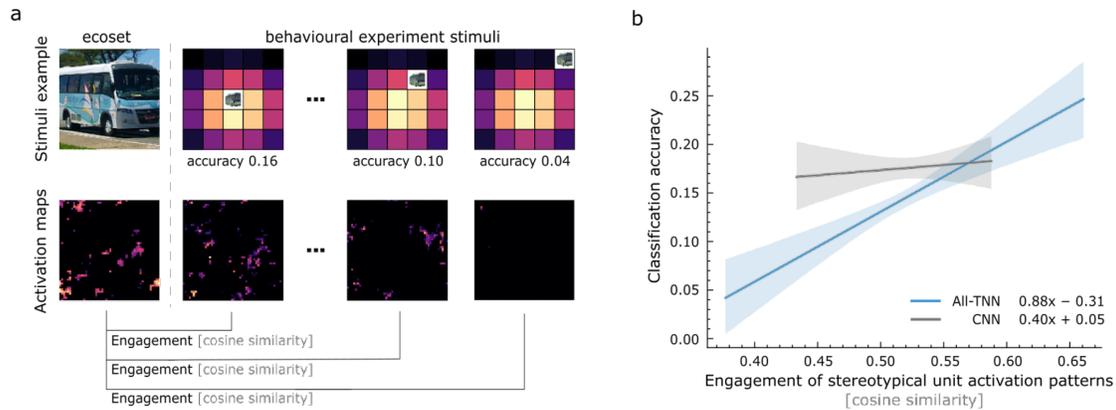

**Figure 7 | Human-like accuracy patterns are linked to topography in All-TNNs. *a*.** *The engagement of stereotypical unit activation patterns by objects presented in various locations was determined by computing a cosine similarity between the corresponding activity maps of the respective network instances (All-TNNs, and CNN controls). In addition, we extracted classification accuracy for all locations. **b**. Relating engagement to accuracy via linear regression with bootstrapping demonstrates that All-TNNs exhibit a positive relationship: stronger engagement of stereotypical unit patterns yield higher classification accuracy. This was not observed for CNNs.*

To investigate this aspect, we recorded average unit activation maps of the final network layer for each object category, based on the ecoset test set. We call these stereotypical activation maps, as they reflect the model's response pattern to images from the distribution of images of a given category on which it was trained. We then recorded the unit activation maps for each stimulus from the behavioural experiment (i.e. each object category, presented in each of the 25 locations of the behavioural experiment; see Methods), and averaged responses across images for each location and category. We call these the experimental unit activation maps, reflecting the model's responses to ouf-of-distribution images used in the behavioural experiment. Comparisons of the stereotypical activation maps to the activation patterns observed during the experimental setting enabled us to test in how far the network's better performance for some locations was due to it engaging the right topographic features. Category-specific activation maps for the stereotypical setting were compared to the 25 experimental locations using cosine similarity (Fig. 7a). If the behavioural effects observed in All-TNNs are driven by their topographical layout, then locations with a better alignment of the stereotypical and experimental activations maps should have a better classification accuracy. Collapsing data across categories and locations, we find that this is indeed the case (linear regression with bootstrapping, slope=0.88; permutation test, n=5e5; *p*<0.001; Fig. 7b). In other words, stimuli are well classified when they successfully engage the right topographic regions of the last layer. In contrast to this, we found that CNNs do not show a significant relationship between unit activation patterns and recognition accuracy (linear regression with bootstrapping, slope=0.40; permutation test, n=5e5; *p*>0.05). These results tie together the two previous results of the paper: the human-like neural topographies of All-TNNs explain their more human-like behavioural patterns.

## Discussion

Here we introduce and test a new artificial neural network architecture, All-TNN, that is capable of modelling topographic aspects of primate vision and their behavioural consequences. All-TNNs fulfill three desiderata for topographical models of the visual system: (1) units with local receptive fields and independently learnt kernels, (2) units arranged on a 2D cortical sheet and (3) spatially smooth feature selectivity. This endows All-TNNs with genuine topography that goes beyond current state-of-the-art CNN models that instead copy and paste identical features across locations.

Using in-silico electrophysiology across the hierarchical levels of All-TNNs, we show that they develop key features of topographic organisation reminiscent of both lower- and higher-level areas of the visual cortex. We find that, unlike non-topographical control networks, All-TNNs tune to spatial occurrence statistics, produce smooth orientation selectivity maps and category-specific topographical organisation. This shows that our simple spatial smoothness constraint is sufficient for All-TNNs trained on natural images to model primate topographies across levels. Our modelling results lend support to the idea that topography in the visual cortex may result from the tendency of spatially neighbouring neurons to learn similar features, which leads to a clustering of neurons. Indeed, the spatial loss that we impose upon the network is in line with experiments that show that neurons preferably connect to neurons with similar orientation selectivity in mammals[46–48].

A surprising property of All-TNNs is a strong centre-periphery topography, with increasingly smooth feature selectivity towards the periphery. This pattern is reminiscent of cortical magnification and could be explained by the trade-off between the classification and spatial loss: affected by dataset statistics, the network learns to detect varied features at the expense of spatial smoothness in the foveal region. The observation that All-TNNs can afford to lose units in the periphery becomes of interest when considering that the brain operates under a limited energy budget and hence likely optimises for energy consumption[49,50]. Our results support the hypothesis that cortical magnification emerged through evolution as an optimal topography to trade off visual performance and neural energy consumption in animals that foveally fixate on relevant objects[39].

Similarly, the finding that All-TNNs show structured, yet varied, representations in the last layer can be tied to the hypothesis that the high-level organisation in IT balances feature variety and homogeneity[24]. Again, the emergence of category-selective clusters can be seen as satisfying a trade-off between the need for varied feature detectors and the tendency to have smooth selectivities.

Our results further reveal a similarity in the developmental principles and trajectory of All-TNNs and the primate brain. The topographic organisation throughout the layers of All-TNN arises early during training and stabilises quickly, while the network continues to increase its classification performance subsequently. This indicates that the topographical organisation offers a stable structure while still allowing for enough flexibility for task learning throughout training. This is similar to the primate visual cortex where early maturation of important architectural structures is thought to provide scaffolding for functional selectivity[23,35,36].

LCN controls, which have an identical architecture to All-TNNs, but lack the spatial similarity loss do not develop topographic features that are similar to the brain, despite being trained on the same dataset and categorization objective. This ties into an important current debate about which aspects of the visual system's structure are genetically hardwired, and which require experience[36,51], and lends support to the idea that both visual expertise (here: dataset) and the right inductive biases (here: architecture and spatial loss) are necessary driving factors for the emergence of functional topographies in the brain. All-TNNs thus invite further modelling of the visual cortex and beyond[52], taking developmental genesis into account through systematic manipulations of architectural features, loss functions, or training datasets to uncover principles and mechanisms underlying the maturation of brain structure and behaviour in the visual system[53].

Importantly, the topographic features of All-TNNs are central to replicating important aspects of human behaviour. Humans can exploit spatial regularities in the typical locations of objects, which may be an adaptive strategy to reduce the computational load and enhance visual efficiency in complex environments[26]. In line with this strategy, neural representations are better decodable and perceptual sensitivity is higher when objects appear at locations that match their typical positions in the world, allowing objects to be more easily detected and recognized when presented in expected locations[26,54]. We find that All-TNNs are closer to human visual behaviour in this setting, due to how their topographical organisation impacts object recognition.

The ability of All-TNNs to link topographies to behaviour allows for new research directions. An obvious hypothesis to test is that if certain objects have more importance than others during training, they may take up more space in the topography, which in turn may account for biases in behaviour[55–57]. As another example, the spatial topography of All-TNNs allows for targeted lesioning of its organised parts, such as the face-selective units in the later layers. This allows using All-TNNs to model brain lesions[58] with potential for clinical impact, and a model of virtual lesioning methods such as Transcranial Magnetic Stimulation (TMS)[59], providing insights into the underlying mechanisms and effects of such experimental interventions.

All-TNNs complement other recent approaches to modelling topography in the visual system[30,41,60–66], which have greatly contributed to our understanding of cortical map formation. One limitation of most existing models of topographic organisation is that they are either truly topographic but not task-performing or task-performing but not truly topographic. Examples of the former are hand-crafted self-organising maps[13,67,68]. The latter are most often, if not always, based on augmenting CNNs, for example by adding a spatial remapping to their units, building self-organising maps based on unit activities, or creating multiple CNN streams[30,60,61,64–66]. While we strongly agree that CNNs can provide important insight into functional organisation, such models do rely on biologically implausible weight sharing rather than genuine topography. All-TNNs are a promising approach to overcome this limitation, as they are both mechanistically truly topographic and task performing.

The current work has several limitations. First, the spatial similarity loss that we used to encourage neighbouring units to learn similar features may not capture the mechanism of smoothness in topographic organization and the exact spatial arrangement of topography in the brain. Future work using All-TNNs as a starting point can explore how smooth maps can emerge naturally from model training, without imposing a secondary spatial similarity constraint explicitly. Possible avenues include using more biologically plausible constraints, such as wiring length optimization[30,64,69], energy constraints[50], cortical size[13,70], or recurrent connectivity patterns. Second, the current set of models is trained on a supervised classification objective, whereas primate visual learning likely relies on unsupervised signals, too[71–73].

In conclusion, All-TNNs are a promising new class of models, which address questions that are beyond the scope of CNNs, and could serve as more accurate models of functional organisation in the visual cortex and its behavioural consequences.

## Methods

### Neural network architectures and training

*All-Topographic Neural Network*

Each layer in All-TNNs is arranged as a 2D sheet, where each unit has its own receptive field and set of weights (i.e., there is no weight sharing, unlike in CNNs). Units in the same "hypercolumn" share the same receptive field. These aspects mirror well-known characteristics of the visual system. In practice, this is implemented by subclassing tensorflow's LocallyConnected2D layer. This layer is arranged in a $height \times width \times channels$ 3D structure, identical to CNNs, but without weight sharing. To convert this 3D LocallyConnectedLayer to our 2D All-TNN layer, we "unfold" each channel to a 2D square, giving rise to a $\sqrt{channels} * height \times \sqrt{channels} * width$ 2D sheet with the desired characteristics.

In addition, each layer has a spatial similarity loss, which promotes similar selectivity for neighbouring units and is crucial for the emergence of topography (as evident in Figure 2). In detail, this spatial similarity loss is computed as the average cosine distance between the weight kernels of neighbouring units in each layer.

The total loss that the model aims to minimize is a composite of two losses: the categorization cross-entropy loss (equation 1), and the spatial similarity loss (summed over all layers; equation 2). The spatial loss is multiplied by a factor α that determines the additive weight of the spatial loss.

(1) $\quad \mathcal{L}_{CE} = -\sum_{c=1}^{M} log(p_c) y_c$

(2) $\quad \mathcal{L}_S = \sum_{l=1}^{L} \frac{1}{N_l} \sum_{n=1}^{N_l} (\alpha_l \cdot \frac{cos\_dist(w_{i,j}, w_{i,j+1}) + cos\_dist(w_{i,j}, w_{i+1,j})}{2})$

(3) $\quad \mathcal{L}_{total} = \mathcal{L}_{CE} + \mathcal{L}_S$

where $w_{i,j}$ is the weight kernel of the unit at the position *i, j* on the 2D sheet, $N_l$ denotes the total number of units in layer *l*, $\alpha_l$ is a hyperparameter in layer *l* that determines the magnitude of the spatial similarity loss, and *L* is the total number of layers in the network..

The All-TNNs we used consist of 6 such layers, of which layers 1, 3, and 5 are followed by 2 by 2 pooling layers. Each layer is subject to L2 regularisation with a factor of 1e-6, and is followed by layer normalisation and a rectified linear unit. We used a spatial loss of $\alpha = 10$ in all layers except the final, which for which we used $\alpha = 10$, due to increased smoothness observed in the higher visual cortex. We used an Adam optimiser with a learning rate of 0.001 and $\epsilon$=0.1. and a regularisation ratio of 1e-6. Weights are initialised with Xavier initialisation. A dropout of 0.2 is applied to all layers during training. See Figure 1 for the specific layer and kernel sizes.

*Locally connected control model*
As a control for the effect of the spatial loss, we also train two All-TNNs with identical hyperparameters but without spatial loss ($\alpha = 0$), meaning that the model trains with only the task loss (cross-entropy).

*Convolutional control model*
Our convolutional controls have the same number of layers, number of units and hyperparameters as our All-TNNs. The spatial similarity loss is not (and cannot be) enforced in this model. It is thus trained with only the cross-entropy loss.

*Dataset & training*
Each model is trained on the ecoset training set (see subsection *Stimuli*). The images were input to the networks with a resolution of 150x150 pixels. Given that "individual differences" exist between ANNs[74], we trained multiple instances of each network type with different random seeds, which are treated as experimental subjects. For All-TNNs and CNNs, we trained 10 instances. For LCNs (i.e. All-TNNs without spatial loss), we only trained two network instances due to resource constraints. All models are trained for 600 epochs.

All models are custom-made, implemented and trained in Python v.3.10 with Tensorflow v.2.8 using NVIDIA A100 GPUs.

**Stimuli**

*Training dataset*
The All-TNN models and CNN control models were trained using the ecoset dataset[34]. The ecoset dataset consists of 1.5 million ecologically motivated images from 565 categories. It was shown that networks trained on ecoset better predict activities in the human higher visual cortex than networks trained on Imagenet (ILSVRC-2012), making it a good choice for modelling the influence of natural image statistics on the emergence of cortical topography.

*Selectivity analysis dataset*
To determine high-level object selectivity in the last layer of the models, we selected 500 images for each superclass used to test selectivity (faces, places and tools). The places and tools images are selected from the 10 most common classes for the respective superclass found in ecoset validation set. The faces are taken from the VGG-Face dataset[75].
*Places*: 'House', 'City', 'Kitchen', 'Mountain', 'Road', 'River', 'Jail', 'Castle', 'Lake', 'Iceberg'
*Tools*: 'Phone', 'Gun', 'Book', 'Table', 'Clock', 'Camera', 'Cup', 'Key', 'Computer', 'Knife'
*Faces*: 10 identities taken from the VGG-Face dataset[75].

*Stimuli for the behavioural experiment*
The spatial classification behaviour of humans, All-TNNs, and control models, was evaluated using segmented objects from the COCO (Common Objects in Context) dataset[43]. This is a large-scale image dataset, gathered from everyday scenes containing common objects, and the objects in this natural sense are precisely labelled, and provided with segmentation masks, bounding box and keypoint annotations.

Easy segmentation of objects from pixel-wise segmentation masks in COCO motivates the use of this dataset for behavioural testing. The stimuli set consists of 16 categories that occur in both ecoset and COCO. These categories are: 'airplane', 'bear', 'broccoli', 'bus', 'cat', 'elephant', 'giraffe', 'kite', 'laptop', 'motorcycle', 'pizza', 'refrigerator', 'scissors', 'toilet', 'train', and 'zebra'. Each category contains 5 exemplars, selected for having similar illumination. To control for visual confounds, all stimuli images were resized to equal size and cropped onto grey backgrounds. Note that we only used 5 stimuli per class in our human experiment due to time limitations, but when we conduct a similar experiment on our networks as described below, we use all COCO stimuli for each class, at each 5x5 location instead.

**Behavioural experiment**

To assess the correspondence between the behaviour of All-TNN, the CNN control model, and human behaviour, we collected human behavioural responses to one image dataset. This allowed for a comparison of the correlation of behavioural responses to the same stimuli between All-TNN, the CNN control model, and human behaviour.

*Participants*
30 healthy adults (aged 21-30 years, mean=25.47 years, SD=2.5 years; 17 female) participated and completed the visual classification task. All participants had normal or corrected-to-normal visual acuity. Prior to participation, ethical approval for the study was obtained to ensure compliance with ethical guidelines. All participants provided written informed consent and received monetary compensation or course credits for their participation.

*Experiment design*
The participants were asked to detect and classify the object that presented on screen. The images used were 215x215 pixels, calculated based on a 5-degree visual angle and a 65 cm screen distance. Stimuli randomly occurred in one of the 25 positions on a 5 by 5 grid on the screen for 40ms. The location of the stimulus was then masked with a Mondrian mask for 300ms. The participants were then presented with a response panel showing the 16 category names after the mask disappeared, after which they had 2150 ms to click on the category name that matched what they saw before the mask. Feedback was given after each trial: the right category name was displayed in a green font colour if the participants were correct, and in red if they were incorrect. The experiment consisted of 2000 trials in total: each object exemplar was shown one time in each location (i.e. for each of the 16 categories, all 5 object exemplars are presented 25 times). The order of stimulus presentation was randomised between participants. Each trial took around 2.5s, and after each set of 200 trials, there was a 2-minute pause. The total experiment took 3 hours in total.

Note that we only used 5 stimuli per class in our human experiment due to time limitations, but when we conduct a similar experiment on our networks as described below, we use all COCO stimuli for each class, at each 5x5 location instead.

*Data recording and processing*
The stimulus presentation in the behavioural experiment was controlled using the Psychtoolbox[76] in Matlab. We recorded the human behavioural data for each object class in each location into 5x5 accuracy maps. We also collected the same data for our models. This involved presenting the models with each object exemplar at each location and recording their classification performance for each position into accuracy maps.

**Data analysis**

*Orientation selectivity*
To determine the orientation selectivity of units in the first layer of the models, we present grating stimuli at 8 angles (equally spaced between 0 and 180 degrees) to the models and record the elicited activity. The gratings have a spatial wavelength of 3 pixels, allowing for >1 cycle within each receptive field in the first layer of our networks. For each grating angle, and for each network unit, we present various phases, and pick the phase that maximizes the unit's activation response to the grating, to find the best alignment between the stimulus and weight kernel. We combine the resulting 8 activities (one per angle) vectorially projecting each angle onto a circle and multiplying it by the corresponding activity, and taking a weighted sum of these vectors (a widespread method for measuring orientation selectivity in electrophysiology[77,78]). Intuitively, if all stimulus angles elicit similar activities, these vectors will cancel out, while there will be a clear winner otherwise. To visualise orientation maps (as well as the entropy and category selectivity maps) for CNNs the kernels are flattened into a 2D-sheet in an identical manner as to the flattening of the All-TNN prior to training.

*Cortical Magnification*
To quantify the diversity of selectivities at each location in the first layer of our models, we calculated entropy in a 3x3 sliding window for each retinotopic position on the orientation selectivity maps. Units that do not respond to any of the grating stimuli are excluded from this analysis. Preferred orientations, computed as described above, are discretized into 8 equally spaced orientation bins, which are used to compute the entropy. This yields a map showing how varied unit responses are (e.g. the entropy is low if all units in the sliding window have similar preferred orientations).

To test if the network can afford to lose units in regions with lower entropy (i.e. more redundant coding), we perform an entropy-based lesion experiment. We use the entropy map described above and lesion the 50% lowest entropy units, and measure the performance of the lesioned network on the validation set of ecoset. As controls, we perform the same test, but lesion the 50% highest entropy units.

*Category selectivity*
We computed the selectivity of each unit in the last layer of our networks to scenes, tools and faces using the d' signal detection measure (see subsection *Stimuli* for the images we used):

$$d' = \frac{|\mu_{in}\mu_{out}|}{mean(\sigma_{in}, \sigma_{out})}$$

where $\mu_{in}$ and $\sigma_{in}$ are the mean and variance of the activations of the unit in response to stimuli of the category of interest, and $\mu_{out}$ and $\sigma_{out}$ are the mean and variance of the activations of the unit in response to stimuli from the other categories. The variances for both distributions are assumed to be in a similar range, therefore we average over the variance of the two distributions in the denominator.

*Positional occurrence maps*
We generated maps that quantify the frequency of occurrence at each image location of each of the 16 COCO categories we use in the behavioural experiment (see Fig. S4a). To do so, we use the binary segmentation masks provided by COCO for each category, and take their average across the whole dataset. These maps are then downsampled to 5x5 using average pooling, to match the dimensionality of accuracy maps derived from behavioural experiments.

*Positional Uncertainty*
Positional uncertainty is derived from the positional occurrence maps described above before downsampling to 5x5. We define positional uncertainty as the area of the locations for which the occurrence frequency is larger than 90% of the highest frequency position (see examples in Fig. S5). A small positional uncertainty thus means that the stimulus instances of an object category often occur in the same positions, while a lard positional uncertainty means that the object appears in varied positions.

*Accuracy Ratio*
The accuracy ratio measures the overall magnitude of positional effects on behavioural performance as the accuracy at the location with the best accuracy divided by the accuracy at the location with the worst accuracy in the accuracy map.

$$accuracy\ ratio = \frac{accuracy_{best}}{accuracy_{worst}}$$

*Accuracy Map Agreement*
Accuracy Map Agreement is a measurement to quantify the alignment of positional dependency in visual behaviour between humans and All-TNNs/CNNs. We computed the Pearson correlation coefficient between the corresponding maps for humans and All-TNNs/CNNs for each of our 16 object categories. The mean noise-ceiling corrected (see below) correlation score is calculated across these 16 categorical accuracy maps using a permutation test.

*Accuracy Dissimilarity Matrix Agreement*

We use Accuracy Dissimilarity Matrices (ADM) to quantify the dissimilarity between pairs of accuracy maps of different categories, in a similar spirit to the well-known Representational Dissimilarity Matrices (RDMs)[45], which compare the dissimilarity between representations of pairs of stimuli pairs of stimuli. Objects yielding behaviourally distinct accuracy maps have high dissimilarity in our ADM, and vice-versa. ADMs were created using Pearson correlation distance between each category accuracy map. To quantify the agreement between model and human ADMs we use noise-ceiling corrected (see below) Spearman correlation between ADMs participants (n=30), and All-TNNs (n=10) or CNNs (n=10) using permutation tests. A higher correlation indicates a higher alignment of the structure of class-wise positional dependencies.

*Noise Ceiling Analysis*

Behaviour is noisy, and therefore, even humans do not correlate perfectly with each other. Therefore, we cannot expect models to have a perfect correlation with our participants either. To account for this, we compute a noise-ceiling by iteratively leaving out one human and seeing how well it correlates with the other 29. This yields 30 correlations, of which we take the mean (this is technically referred to as the noise ceiling lower bound). We then divide our model-human correlations by this value.

*Stereotypical unit activation patterns and link to behaviour*

To ask whether the drop in performance at different locations of the 5x5 accuracy map can be explained by the fact that All-TNNs fail to engage the adequate unit population, we extracted activation maps from the last layer of All-TNNs and CNN control models in response to both ecoset test set images for our 16 classes, and stimuli from COCO displayed on the 5x5 experimental grid. We consider the responses to ecoset test images "stereotypical", and quantify the engagement of these stereotypical unit activation patterns for each class and location in the experimental 5x5 grid. In detail, for each of our 16 classes, we compute the cosine similarity between the "stereotypical" activation map and the "experimental" activation map elicited by presenting our experimental stimuli at each location. We then relate this stereotypical engagement to classification accuracy by using a linear regression with bootstrapping.

**Code and data availability**

All analyses of human and model data were performed in custom Python software, making use of Numpy and/or scikit-learn packages. The code and data required to reproduce our results will be released upon journal publication of this paper.

**Acknowledgements**

The authors acknowledge support by the following grants: A.D. is supported by SNF grant n.203018. T.C.K., V.B., and A.D. are supported by the ERC STG grant 101039524 TIME. D.K. is supported by the Deutsche Forschungsgemeinschaft (SFB/TRR 135, project number 222641018), an ERC Starting Grant (ERC-2022-STG 101076057), and by "The Adaptive Mind," funded by the Excellence Program of the Hessian Ministry of Higher Education,

Science, Research and Art. Z.L. is supported by CSC grant (202106120015) and R.M.C. is supported by DFG grants CI241/1-1, CI241/3-1, CI241/7- 1 and ERC STG grant 803370.

**Bibliography**

1. Doerig, A. *et al.* The neuroconnectionist research programme. *Nat. Rev. Neurosci.* 1–20 (2023) doi:10.1038/s41583-023-00705-w.

2. Cichy, R. M. & Kaiser, D. Deep Neural Networks as Scientific Models. *Trends Cogn. Sci.* **23**, 305–317 (2019).

3. Richards, B. A. *et al.* A deep learning framework for neuroscience. *Nat. Neurosci.* **22**, 1761–1770 (2019).

4. Khaligh-Razavi, S.-M. & Kriegeskorte, N. Deep Supervised, but Not Unsupervised, Models May Explain IT Cortical Representation. *PLoS Comput. Biol.* **10**, e1003915 (2014).

5. Yamins, D., Hong, H., Cadieu, C. & Dicarlo, J. J. Hierarchical Modular Optimization of Convolutional Networks Achieves Representations Similar to Macaque IT and Human Ventral Stream. *Adv. Neural Inf. Process. Syst. NIPS* (2013).

6. Yamins, D. L. K. *et al.* Performance-optimized hierarchical models predict neural responses in higher visual cortex. *Proc. Natl. Acad. Sci.* **111**, 8619–8624 (2014).

7. Güçlü, U. & Gerven, M. A. J. van. Deep Neural Networks Reveal a Gradient in the Complexity of Neural Representations across the Ventral Stream. *J. Neurosci.* **35**, 10005–10014 (2015).

8. Kell, A. J. E., Yamins, D. L. K., Shook, E. N., Norman-Haignere, S. V. & McDermott, J. H. A Task-Optimized Neural Network Replicates Human Auditory Behavior, Predicts Brain Responses, and Reveals a Cortical Processing Hierarchy. *Neuron* **98**, 630-644.e16 (2018).

9. Spoerer, C. J., Kietzmann, T. C., Mehrer, J., Charest, I. & Kriegeskorte, N. Recurrent neural networks can explain flexible trading of speed and accuracy in biological vision. *PLOS Comput. Biol.* **16**, e1008215 (2020).

10. Kar, K., Kubilius, J., Schmidt, K., Issa, E. B. & DiCarlo, J. J. Evidence that recurrent circuits are critical to the ventral stream's execution of core object recognition behavior. *Nat. Neurosci.* **22**, 974–983 (2019).

11. Rust, N. C. & Mehrpour, V. Understanding Image Memorability. *Trends Cogn. Sci.* **24**, 557–568 (2020).

12. Kaschube, M. *et al.* Universality in the Evolution of Orientation Columns in the Visual Cortex. *Science* **330**, 1113–1116 (2010).

13. Najafian, S. *et al.* A theory of cortical map formation in the visual brain. *Nat. Commun.* **13**, 2303 (2022).

14. Hubel, D. H. & Wiesel, T. N. Receptive fields, binocular interaction and functional architecture in the cat's visual cortex. *J. Physiol.* **160**, 106–154 (1962).

15. Jung, Y. J. *et al.* Orientation pinwheels in primary visual cortex of a highly visual marsupial. *Sci. Adv.* **8**, eabn0954 (2022).

16. Hubel, D. H. & Wiesel, T. N. Ferrier lecture - Functional architecture of macaque monkey visual cortex. *Proc. R. Soc. Lond. B Biol. Sci.* **198**, 1–59 (1997).

17. Kanwisher, N., McDermott, J. & Chun, M. M. The Fusiform Face Area: A Module in Human Extrastriate Cortex Specialized for Face Perception. *J. Neurosci.* **17**, 4302–4311 (1997).

18. Tsao, D. Y., Freiwald, W. A., Tootell, R. B. H. & Livingstone, M. S. A cortical region consisting entirely of face-selective cells. *Science* **311**, 670–674 (2006).

19. Peelen, M. V. & Downing, P. E. Selectivity for the human body in the fusiform gyrus. *J. Neurophysiol.* **93**, 603–608 (2005).

20. Dilks, D. D., Julian, J. B., Paunov, A. M. & Kanwisher, N. The Occipital Place Area Is Causally and Selectively Involved in Scene Perception. *J. Neurosci.* **33**, 1331–1336 (2013).

21. Nasr, S. *et al.* Scene-Selective Cortical Regions in Human and Nonhuman Primates. *J. Neurosci.* **31**, 13771–13785 (2011).

22. Epstein, R. & Kanwisher, N. A cortical representation of the local visual environment. *Nature* **392**, 598–601 (1998).

23. Deen, B. *et al.* Organization of high-level visual cortex in human infants. *Nat. Commun.* **8**, 13995 (2017).

24. Tanaka, K. Columns for complex visual object features in the inferotemporal cortex: clustering of cells with similar but slightly different stimulus selectivities. *Cereb. Cortex N. Y. N 1991* **13**, 90–99 (2003).

25. Konkle, T. & Caramazza, A. Tripartite Organization of the Ventral Stream by Animacy and Object Size. *J. Neurosci.* **33**, 10235–10242 (2013).

26. Kaiser, D., Quek, G. L., Cichy, R. M. & Peelen, M. V. Object Vision in a Structured World. *Trends Cogn. Sci.* **23**, 672–685 (2019).

27. Kaiser, D. & Cichy, R. M. Typical visual-field locations enhance processing in object-selective channels of human occipital cortex. *J. Neurophysiol.* **120**, 848–853 (2018).

28. Bar, M. Visual objects in context. *Nat. Rev. Neurosci.* **5**, 617–629 (2004).

29. Song, H. F., Kennedy, H. & Wang, X.-J. Spatial embedding of structural similarity in the cerebral cortex. *Proc. Natl. Acad. Sci.* **111**, 16580–16585 (2014).

30. Blauch, N. M., Behrmann, M. & Plaut, D. C. A connectivity-constrained computational account of topographic organization in primate high-level visual cortex. *Proc. Natl. Acad. Sci.* **119**, e2112566119 (2022).

31. Finzi, D. *et al.* Differential spatial computations in ventral and lateral face-selective regions are scaffolded by structural connections. *Nat. Commun.* **12**, 2278 (2021).

32. Dumoulin, S. O. & Wandell, B. A. Population receptive field estimates in human visual cortex. *NeuroImage* **39**, 647–660 (2008).

33. Aflalo, T. N. & Graziano, M. S. A. Organization of the Macaque Extrastriate Visual Cortex Re-Examined Using the Principle of Spatial Continuity of Function. *J. Neurophysiol.* **105**, 305–320 (2011).

34. Mehrer, J., Spoerer, C. J., Jones, E. C., Kriegeskorte, N. & Kietzmann, T. C. An ecologically motivated image dataset for deep learning yields better models of human vision. *Proc. Natl. Acad. Sci.* **118**, e2011417118 (2021).

35. Ellis, C. T. *et al.* Retinotopic organization of visual cortex in human infants. *Neuron* **109**, 2616-2626.e6 (2021).

36. Arcaro, M. J., Schade, P. F., Vincent, J. L., Ponce, C. R. & Livingstone, M. S. Seeing faces is necessary for face-patch formation. *Nat. Neurosci.* **20**, 1404–1412 (2017).

37. Cowey, A. & Rolls, E. T. Human cortical magnification factor and its relation to visual acuity. *Exp. Brain Res.* **21**, 447–454 (1974).

38. Levi, D. M., Klein, S. A. & Aitsebaomo, A. P. Vernier acuity, crowding and cortical magnification. *Vision Res.* **25**, 963–977 (1985).

39. Provis, J. M., Dubis, A. M., Maddess, T. & Carroll, J. Adaptation of the Central Retina for High Acuity Vision: Cones, the Fovea and the Avascular Zone. *Prog. Retin. Eye Res.* **35**, 63–81 (2013).

40. Kanwisher, N. Functional specificity in the human brain: A window into the functional architecture of the mind. *Proc. Natl. Acad. Sci.* **107**, 11163–11170 (2010).

41. Bao, P., She, L., McGill, M. & Tsao, D. Y. A map of object space in primate inferotemporal cortex. *Nature* **583**, 103–108 (2020).

42. de Haas, B. *et al.* Perception and Processing of Faces in the Human Brain Is Tuned to Typical Feature Locations. *J. Neurosci. Off. J. Soc. Neurosci.* **36**, 9289–9302 (2016).

43. Lin, T.-Y. *et al.* Microsoft COCO: Common Objects in Context. Preprint at http://arxiv.org/abs/1405.0312 (2015).

44. Fahrenfort, J. J., Scholte, H. S. & Lamme, V. a. F. Masking disrupts reentrant processing in human visual cortex. *J. Cogn. Neurosci.* **19**, 1488–1497 (2007).

45. Nili, H. *et al.* A Toolbox for Representational Similarity Analysis. *PLOS Comput. Biol.* **10**, e1003553 (2014).

46. Malach, R., Amir, Y., Harel, M. & Grinvald, A. Relationship between intrinsic connections and functional architecture revealed by optical imaging and in vivo targeted biocytin injections in primate striate cortex. *Proc. Natl. Acad. Sci. U. S. A.* **90**, 10469–10473 (1993).

47. Gilbert, C. D. & Wiesel, T. N. Columnar specificity of intrinsic horizontal and corticocortical connections in cat visual cortex. *J. Neurosci. Off. J. Soc. Neurosci.* **9**, 2432–2442 (1989).

48. Fitzpatrick, D. The Functional Organization of Local Circuits in Visual Cortex: Insights from the Study of Tree Shrew Striate Cortex. *Cereb. Cortex* **6**, 329–341 (1996).

49. Olshausen, B. A. & Field, D. J. Emergence of simple-cell receptive field properties by learning a sparse code for natural images. *Nature* **381**, 607–609 (1996).

50. Ali, A., Ahmad, N., Groot, E. de, Gerven, M. A. J. van & Kietzmann, T. C. Predictive coding is a consequence of energy efficiency in recurrent neural networks. *Patterns* **3**, (2022).

51. Ratan Murty, N. A. *et al.* Visual experience is not necessary for the development of face-selectivity in the lateral fusiform gyrus. *Proc. Natl. Acad. Sci.* **117**, 23011–23020 (2020).

52. Eggermont, J. J. The Role of Sound in Adult and Developmental Auditory Cortical Plasticity. *Ear Hear.* **29**, 819–829 (2008).

53. Ibbotson, M. & Jung, Y. J. Origins of Functional Organization in the Visual Cortex. *Front. Syst. Neurosci.* **14**, 10 (2020).

54. Kaiser, D. & Cichy, R. M. Typical visual-field locations facilitate access to awareness for everyday objects. *Cognition* **180**, 118–122 (2018).

55. Mahon, B. Z. & Caramazza, A. What drives the organization of object knowledge in the brain? The distributed domain-specific hypothesis. *Trends Cogn. Sci.* **15**, 97–103 (2011).

56. Bonner, M. F. & Epstein, R. A. Coding of navigational affordances in the human visual system. *Proc. Natl. Acad. Sci.* **114**, 4793–4798 (2017).

57. Op de Beeck, H. P., Pillet, I. & Ritchie, J. B. Factors Determining Where Category-Selective Areas Emerge in Visual Cortex. *Trends Cogn. Sci.* **23**, 784–797 (2019).

58. Liu, T. T. & Behrmann, M. Functional outcomes following lesions in visual cortex: Implications for plasticity of high-level vision. *Neuropsychologia* **105**, 197–214 (2017).

59. Silvanto, J. & Cattaneo, Z. Common framework for "virtual lesion" and state-dependent TMS: The facilitatory/suppressive range model of online TMS effects on behavior. *Brain Cogn.* **119**, 32–38 (2017).

60. Doshi, F. R. & Konkle, T. Cortical topographic motifs emerge in a self-organized map of object space. *Sci. Adv.* **9**, eade8187 (2023).

61. Margalit, E. *et al.* A Unifying Principle for the Functional Organization of Visual Cortex. 2023.05.18.541361 Preprint at https://doi.org/10.1101/2023.05.18.541361 (2023).

62. Kanwisher, N., Gupta, P. & Dobs, K. CNNs reveal the computational implausibility of the expertise hypothesis. *iScience* **26**, 105976 (2023).

63. Keller, T. A. & Welling, M. Topographic VAEs learn Equivariant Capsules. in *Advances in Neural Information Processing Systems* vol. 34 28585–28597 (Curran Associates, Inc., 2021).

64. Zhang, Y., Zhou, K., Bao, P. & Liu, J. Principles governing the topological organization of object selectivities in ventral temporal cortex. 2021.09.15.460220 Preprint at https://doi.org/10.1101/2021.09.15.460220 (2021).

65. Dobs, K., Martinez, J., Kell, A. J. E. & Kanwisher, N. Brain-like functional specialization emerges spontaneously in deep neural networks. *Sci. Adv.* **8**, eabl8913 (2022).

66. Lindsey, J., Ocko, S. A., Ganguli, S. & Deny, S. A Unified Theory of Early Visual Representations from Retina to Cortex through Anatomically Constrained Deep CNNs. *arXiv* (2019) doi:10.48550/arxiv.1901.00945.

67. Kohonen, T. Self-organized formation of topologically correct feature maps. *Biol. Cybern.* **43**, 59–69 (1982).

68. Swindale, N. V. & Bauer, H.-U. Application of Kohonen's self-organizing feature map algorithm to cortical maps of orientation and direction preference. *Proc. R. Soc. B Biol. Sci.* **265**, 827–838 (1998).

69. Koulakov, A. A. & Chklovskii, D. B. Orientation Preference Patterns in Mammalian Visual Cortex A Wire Length Minimization Approach. *Neuron* **29**, 519–527 (2001).

70. Weigand, M., Sartori, F. & Cuntz, H. Universal transition from unstructured to structured neural maps. *Proc. Natl. Acad. Sci.* **114**, E4057–E4064 (2017).

71. Zhuang, C. *et al.* Unsupervised neural network models of the ventral visual stream. *Proc. Natl. Acad. Sci.* **118**, (2021).

72. Konkle, T. & Alvarez, G. A. A self-supervised domain-general learning framework for human ventral stream representation. *Nat. Commun.* **13**, 491 (2022).

73. Storrs, K. R., Anderson, B. L. & Fleming, R. W. Unsupervised learning predicts human perception and misperception of gloss. *Nat. Hum. Behav.* **5**, 1402–1417 (2021).

74. Mehrer, J., Spoerer, C. J., Kriegeskorte, N. & Kietzmann, T. C. Individual differences among deep neural network models. *Nat. Commun.* **11**, 5725 (2020).

75. Parkhi, O. M., Vedaldi, A. & Zisserman, A. Deep Face Recognition. in *Procedings of the British Machine Vision Conference 2015* 41.1-41.12 (British Machine Vision Association, 2015). doi:10.5244/C.29.41.

76. Brainard, D. H. The Psychophysics Toolbox. *Spat. Vis.* **10**, 433–436 (1997).

77. Hübener, M., Shoham, D., Grinvald, A. & Bonhoeffer, T. Spatial Relationships among Three Columnar Systems in Cat Area 17. *J. Neurosci. Off. J. Soc. Neurosci.* **17**, 9270–84 (1998).

78. Kaschube, M., Schnabel, M. & Wolf, F. Self-organization and the selection of pinwheel density in visual cortical development. *New J. Phys.* **10**, 015009 (2008).

# Supplementary Material

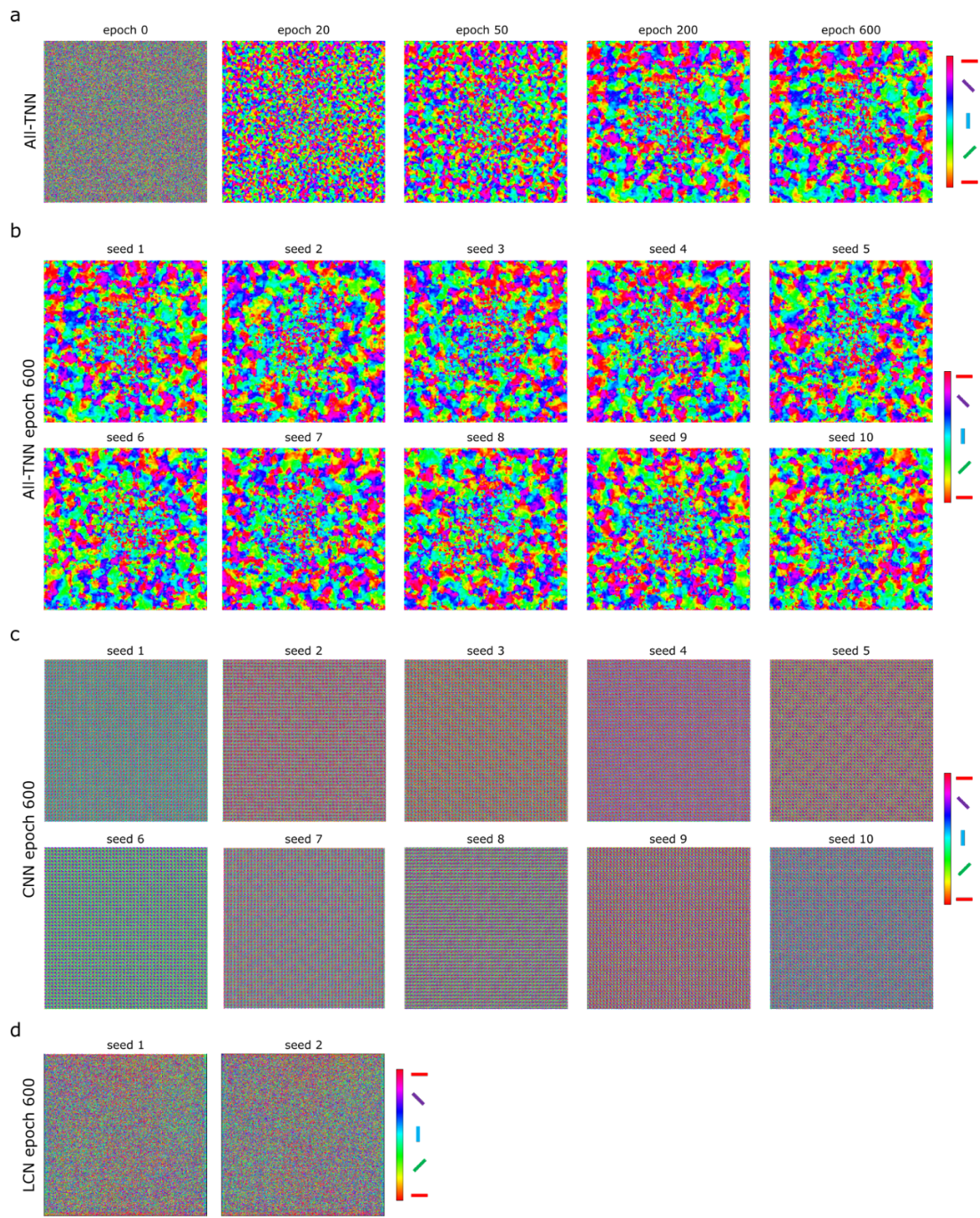

**Figure S1 | Orientation selectivity maps across training epochs and model instances.**
*a. The V1-like organisation of orientation selectivities in the first layer of All-TNN remains stable after emergence in the first training epochs. b. The organisation of orientation selectivities for All-TNNs is consistent across all trained network seeds. c. Unstructured salt-and-pepper orientation selectivity maps emerge in all trained CNN seeds. d. Salt-and-pepper orientation selectivity maps emerge in all trained LCN seeds.*

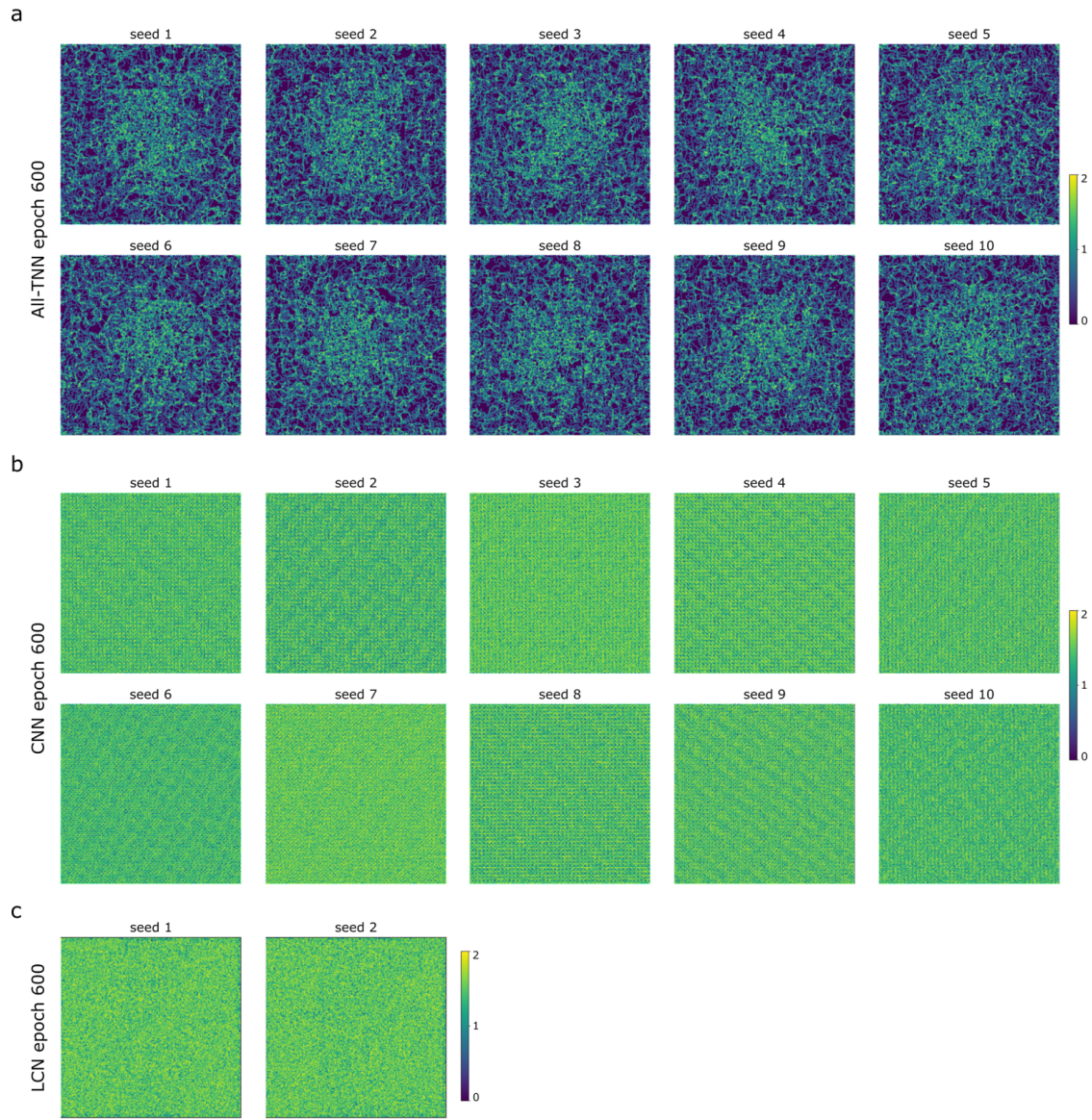

**Figure S2 | Entropy analysis across model instances.** *a. The cortical magnification of entropy maps in the first layer of All-TNN emerges consistently across network seeds. b. Homogenous unstructured entropy maps emerge in all trained CNN seeds. c. Homogenousentropy maps emerge in all trained LCN seeds.*

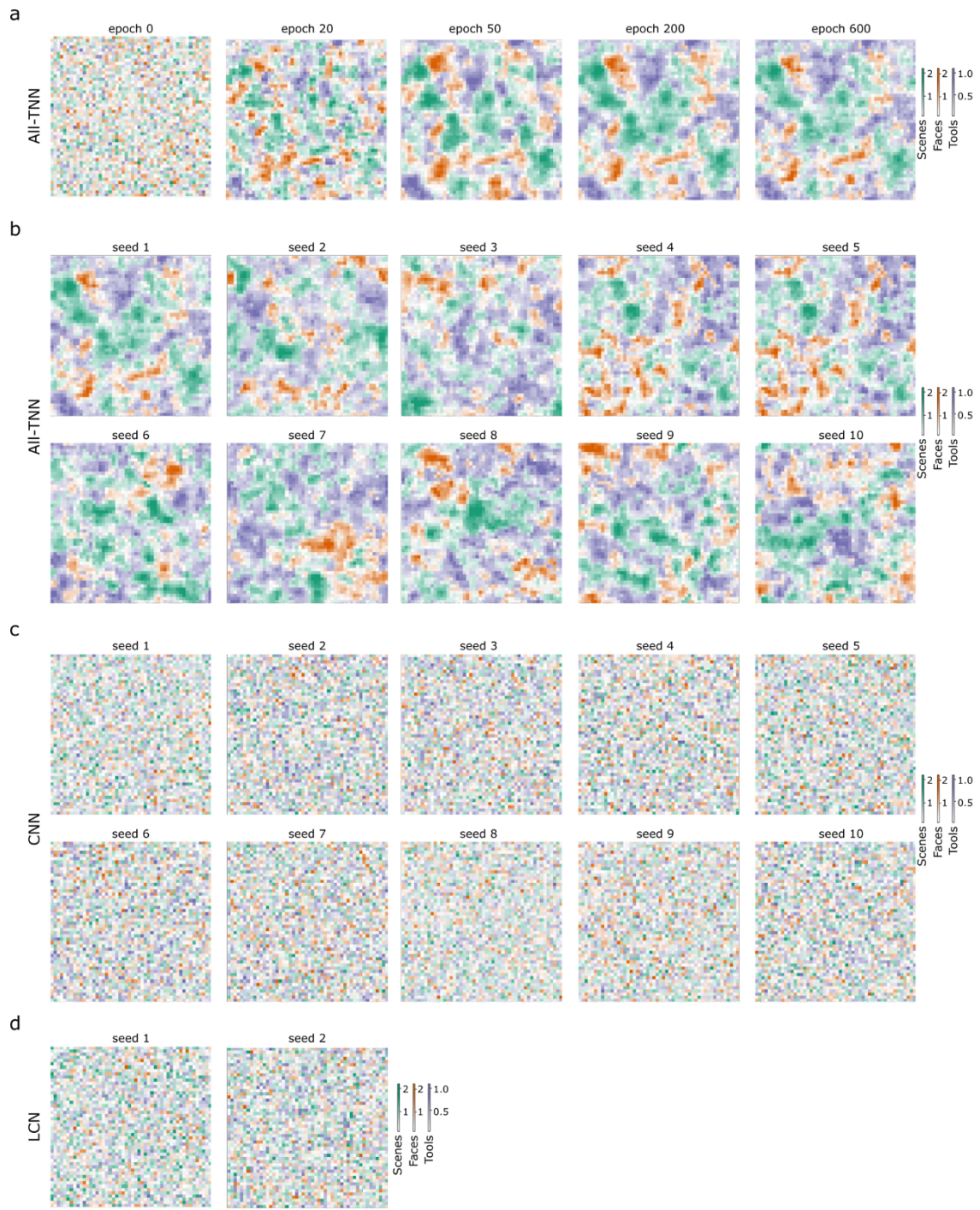

**Figure S3 | Category selectivity maps across training epochs and model instances.** *a. The clustering of high-level category-based selectivities (d') for tools, scenes, and faces in the last layer of All-TNN emerges through training epochs. b. The emergence of clusters of category selectivity for All-TNNs is consistent across all trained network seeds. c. Category selectivity maps are unstructured in all trained CNN seeds d. Category selectivity maps are unstructured in all trained LCN seeds.*

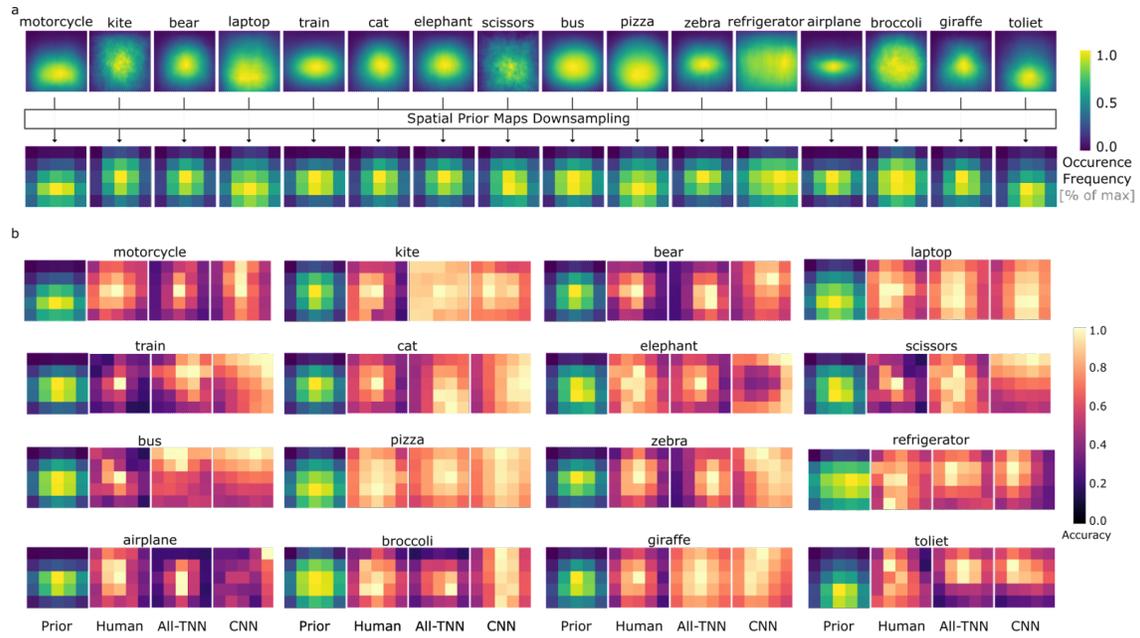

**Figure S4 | Categorical spatial prior and accuracy maps** *a. Spatial occurrence frequency maps for different object categories in high resolution, derived from the COCO dataset, are downsampled to 5x5 resolution to match the resolution of accuracy map results from behaviour experiments.* ***b.*** *Visualisation of accuracy maps for all 16 categories for average humans, All-TNNs and CNNs.*

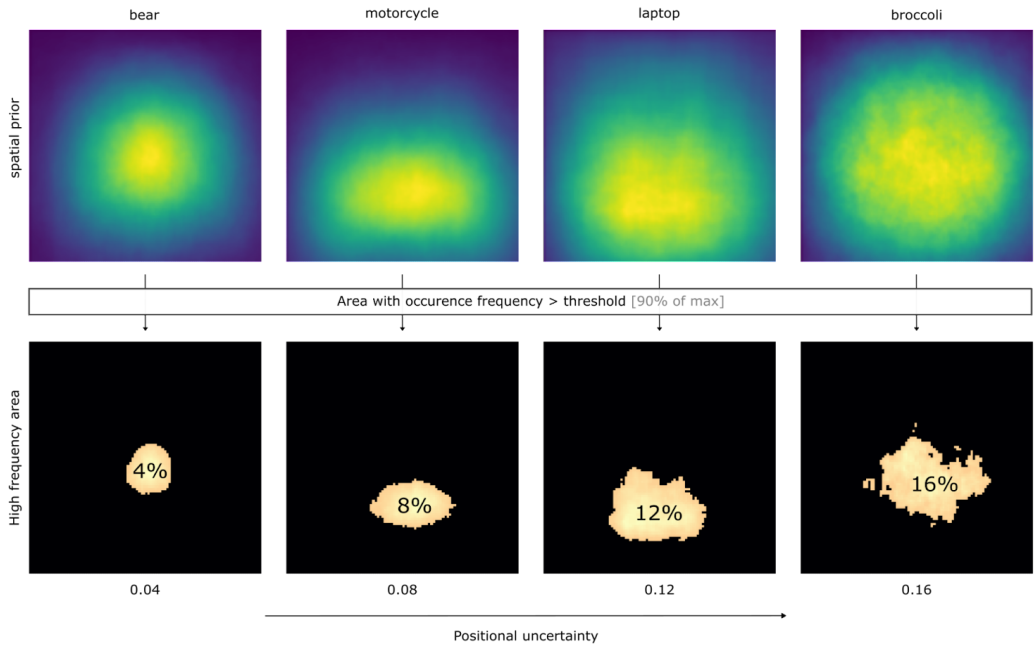

**Figure S5 | Positional uncertainty.** The calculation of categorical positional uncertainty is based on the area size of the locations with the occurrence frequency larger than threshold (90% of highest frequency) on the categorical positional occurrence maps. The size of the region with a high frequency of occurrence represents the magnitude of the categorical position uncertainty.